\documentclass[twocolumn,superscriptaddress,prl,showpacs]{revtex4}
\usepackage[letterpaper,left=0.7in,right=0.7in,top=0.90in,bottom=0.8in]{geometry}
\usepackage{amssymb}
\usepackage{amsmath}
\usepackage{graphicx}
\usepackage{subfigure}

\newcommand{\qsgw}{QS\textit{GW}}

\begin{document}

\title{
Effect of electron correlations  on  (001) Fe/MgO interfaces.
}

\author{Sergey V. Faleev}
\email{sfaleev@nantworks.com}
\author{Oleg N. Mryasov}
\email{omryasov@mint.ua.edu}
\affiliation{Physics and Astronomy, University of Alabama, Tuscaloosa, AL, 35487, USA}
\author{Mark van Schilfgaarde}
\affiliation{Arizona State University, Tempe, Arizona, 85287, USA}

\date{\today}

\begin{abstract}
We developed a parametrization of transmission probability that reliably captures essential elements of the tunneling process in magnetic tunnel junctions.
The electronic structure  of  Fe/MgO system is calculated  within the quasiparticle self-consistent \textit{GW} approximation and used
to evaluate transmission probability across (001) Fe/MgO interface.
The transmission
has a peak at +0.12~V, in excellent agreement with recent differential conductance measurements for electrodes with antiparallel spin.
These findings confirm that the observed current-voltage characteristics
are intrinsic to well defined (001) Fe/MgO interfaces, in contrast to previous
predictions based on the local spin-density approximation, and also that
many-body effects are important to realistically describe electron
transport across  well defined metal-insulator interfaces.
\end{abstract}
\maketitle


Energy dissipation in switching a logic unit (transistor) is perhaps the most important
bottleneck to continued realization of Moore's law scaling in the density
of integrated circuits \cite{roadmap,solomon}. New materials with special,
well defined interfaces \cite{schlom} and functional properties
\cite{hwang} offer promising routes to circumvent losses.
Most new schemes exploit tunneling phenomena  across metal-insulators or semiconductor interfaces, so they will play an increasingly important role as we explore fundamental limits of density (miniaturization) in integrated circuits \cite{solomon}.

In thin FM/insulator/FM heterostructures, where an FM is a
ferromagnet, spin polarized electron tunneling is observed.  In such
``magnetic tunnel junctions'' (MTJs) the tunneling resistance changes
when alignment of the two FM electrodes are switched from an
anti-parallel configuration (APC) to a parallel configuration (PC).
This property is encapsulated in the tunneling magnetoresistance
(TMR), $T^{MR}=(G_P - G_A)/G_A$.  $G_P$ and $G_A$ are conductivities
in the PC and APC~\cite{Zhang03,Tsymbal03}.  These heterostructures
are of particular technological interest because of the recent
discovery of very large TMR in highly crystalline (001) Fe/MgO/Fe
MTJs \cite{Yuasa04,Parkin04}.  Equally important to the large TMR is
the record low critical current needed to manipulate (switch)
magnetization without external magnetic field, relative ease of
fabrication and reproducibility \cite{Yuasa07}. At a fundamental
level, the (001) Fe/MgO/Fe system presents an excellent opportunity to
investigate the role of the interface and test our ability to predict
critical properties controlling transport in
a heterostructure with well defined interfaces \cite{Yuasa04}.

The local spin density approximation (LSDA) has been used to predict
TMR in Fe/MgO/Fe system
\cite{Butler01,Belashchenko05,Rungger09,Tiusan04,Feng09}.  Reports are
largely in agreement with each other but at variance with differential
conductance measurements for thin MTJs \cite{Zermatten08,Du10}.  The
LSDA predicts a narrow band of interface resonance states (IRS) of
minority electrons (Fig. 1) which overlaps the Fermi level $E_F$.
This prediction has two consequences: first, a sharp reduction in TMR
at voltages on the order of $\sim$0.02 V, due to sharp (resonant)
reduction in minority spin contribution to differential conductance in
PC \cite{Rungger09,Du10}.  It is called the `zero-bias
anomaly'. Secondly, it predicts a rise in TMR at larger voltage,
$\sim$0.1 V, due to reduction in differential conductance in APC
\cite{Rungger09}.  Neither effect has been observed experimentally;
instead differential conductance in APC increases monotonically, with
$d^2I/dV^2$ reaching a peak near 0.12 V \cite{Zermatten08} or 0.15 V
\cite{Du10}.  Zermatten et al.  attributed this effect to interface
states, observed at the (001) Fe free surface by STM measurements.

The absence of `zero-bias anomaly' in PC is usually explained as a
slight asymmetry of the electrodes from e.g., disorder, which breaks
matching of the resonant states.  Thus for any applied bias the
minority-spin contribution to current in PC is much smaller than the
majority-spin contribution for MgO thicknesses larger than 1nm
\cite{Belashchenko05}.  As regards the APC, whether inconsistencies
between the LSDA and experiment are due to some extrinsic phenomenon
(e.g. disorder), or a failure in the LSDA, has remained an open
question.
The LSDA underestimates bandgaps, which strongly affects
the 
band structure at imaginary $k$ governing evanescent decay.  It also
poorly describes the
Schottky barrier height \cite{Das89}.  Thus in the LSDA, most of the key
parameters responsible for TMR  are somewhat suspect.




Here we investigate electronic structure of (001) Fe/MgO/Fe using the
Quasiparticle Self-consistent \textit{GW} (\qsgw) approximation for the
electronic structure, which does not rely on the LSDA. We find significant
corrections to energy and $k$-space character of the minority spin channel
interface states.  We investigate how these electronic structure
corrections alter I-V characteristics in the APC, and account for available
experimental results \cite{Zermatten08,Du10}.
\qsgw\ uses self-consistency to minimize the many-body part of the
hamiltonian, thus allowing accurate determination of quasiparticle (QP)
levels with low-order diagrams.  It has been tested for a wide variety of
bulk material systems and has been shown to be a good predictor of
materials properties for many classes of compounds composed of elements
throughout the periodic table \cite{Faleev04,Schilfgaarde06,Kotani07}.  It
vastly improves on the accuracy of any existing DFT method, as well as
standard implementations of $GW$, namely (1-shot) perturbations around the
LSDA.  Also it surmounts problems of self-consistency inherent in a true
self-consistent \emph{GW} scheme.  Self-consistency is particularly
important in ionic compounds such as MgO \cite{Schilfgaarde06}.  Moreover,
it may be necessary for reliable description of the metal/insulator
interface since screening at the interface can be particularly important.
For example, screening has been shown to strongly affect the molecular
levels of benzene near a graphite interface \cite{Louie06}, a correlation
effect not captured by a Kohn-Sham theory.  As we show here, \qsgw\ applied
to this highly heterogeneous system appears to have the same uniform
accuracy found in homogeneous materials systems.

We develop a formula to obtain the transmission probability $T$ that
avoids direct calculation of the tunneling via Landauer-Buttiker theory.
$T$ is parameterized by the local density of states (DOS) inside the
tunneling layer.  As we will show, it does an excellent job at reproducing
the full Landauer-Buttiker transmission for a given one-body hamiltonian;
it thus makes possible predictions of transport within the \qsgw\
approximation.
Used in conjunction with \qsgw, $T$ calculated in APC is in excellent
agreement with observed I-V characteristics.  This demonstrates that the
observed TMR is not an artifact of imperfections at the interface, but an
\emph{intrinsic property} of it.  An important corollary is that many-body
corrections to the LSDA for electronic structure can have profound effect
on transport properties.  In this particular case the dominant correction
to the LSDA is a shift in the IRS, of the same magnitude as a typical bias
voltage.  But generally speaking we can expect \qsgw\ to describe
electronic structure in inhomogeneous systems with vastly better accuracy
than commonly adopted approaches and thus investigate implication of these
corrections for transport.


Owing to heavy computational costs, the Fe/MgO interface is modeled in
\qsgw\ with a periodic slab of 5 Fe and 5 MgO layers ordered on the (001)
plane.
All results reported here adopt a
generalized linear muffin-tin orbitals (LMTO) method
\cite{Schilfgaarde06,Kotani07}, and use the relaxed nuclear coordinates of
the Fe/MgO interface from Ref.~\cite{Worthmann04}.

\begin{figure}[t]
\includegraphics*[width=8.5cm]{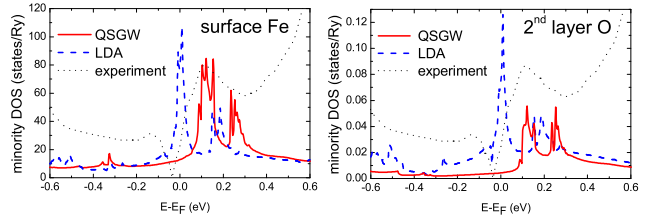} 
\caption{(color online).  Minority electron DOS projected to the surface Fe
layer (left panel) and the oxygen atom in the second layer from the Fe/MgO
interface (right panel).  \qsgw\ results are depicted by a (red) solid
line, LSDA by a (blue) dashed line.  Note that the panels have different
scales. Black dotted line depicts $|d^2I/dV^2|$ measured in the
antiparallel configuration \cite{Zermatten08} (arbitrary units).  }
\label{DOS}
\end{figure}

Minority DOS of the Fe/MgO superlattice, projected on the Fe interfacial
atom, and onto the oxygen atom in the second layer from the interface, are
depicted in Fig.~\ref{DOS}.  For both LSDA and \qsgw\ two narrow peaks are
seen, separated by $\sim$0.15 eV. 
The iron and oxygen projected DOS have similar shapes,
indicating that these peaks originate from the same interface resonance
states.  These states fall near midgap in MgO and decay exponentially,
the decay being more pronounced in \qsgw.

The LSDA puts one peak just at $E_F$, in agreement with several prior LSDA
calculations \cite{Butler01,Belashchenko05,Rungger09,Tiusan04,Feng09}).
There is a corresponding IRS resonance in \qsgw, but it falls at
$E_F$+0.12~eV.  This difference is important: it is comparable to typical bias
voltages.


Through calculation of $T$ we can establish a connection with I-V
measurements (black dotted line in Fig.~\ref{DOS}).  The observed
$|d^2I/dV^2|$ has peaks at both +0.12 V and $-$0.12~V, with the former
being more pronounced.  The difference can be explained by the structural
asymmetry \cite{Zermatten08}: the top interface is grown last and is
rougher than the bottom one.  In the following we show quantitatively that
a combination of featureless DOS in the majority channel (not shown) and
sharply peaked IRS states in the minority channel explains the observed
peak in the differential conductance $|d^2I/dV^2|$ at +0.12 V. (Voltage is
defined so that forward bias samples minority states of the bottom
electrode with $E{>}E_F$).  We use the following
expression for $T$ at zero applied bias:
\begin{equation}
T^D_{\sigma \sigma^\prime}(E) = \frac{\lambda}{N_{||}}\sum_{\mathbf{k}} {\rm
DOS}^{\sigma}_{\mathbf{k} E}(L)\, e^{-2\gamma_{\mathbf{k}E}d_{LR}}
\, {\rm DOS}^{\sigma^\prime}_{\mathbf{k} E}(R)
\label{TDD}
\end{equation}
$N_{||}$ is the number of \textbf{k}-points in the 2D Brillouin zone normal
to the interface, $\sigma$ and $\sigma^\prime$ denote spin polarizations of
the left and right electrodes, ${\rm DOS}^{\sigma}_{\mathbf{k} E}(L)$ is
the DOS of electrons with spin $\sigma$ projected onto nucleus $L$, chosen
at will somewhere in the MgO close to the left Fe/MgO interface, and ${\rm
  DOS}^{\sigma'}_{\mathbf{k} E}(R)$ is the corresponding DOS of spin
$\sigma^\prime$, projected onto a nucleus $R$ in MgO close to the right
Fe/MgO interface.  $\gamma_{\mathbf{k}E}$ is the smallest
(spin-independent) imaginary wave number for evanescent states inside the
MgO barrier with given $\mathbf{k}$ and energy $E$--a property of the
complex band structure of bulk MgO.  $d_{LR}$ is the spacing between planes
containing atoms $L$ and $R$.  Eqn.~(\ref{TDD}) neglects parallel channels
with larger imaginary wave number, which is always satisfied if the barrier
is thick enough.  It becomes exact for one-dimensional case if the
projected DOS is replaced by the square of the wave function $\psi$
propagating in corresponding electrode that is normalized to carry unit
flux \cite{LandauQM,Bel04}.
Since we replace the flux-normalized $|\psi|^2$ with a
local DOS, we include factor $\lambda$ to correct for the
(effectively unnormalized) $\psi$.
Once local DOS and $\gamma_{\mathbf{k}E}$ are given, $T$ can be calculated
for \emph{arbitrary} MgO thickness.  Though Eq.~(\ref{TDD}) bears a
superficial resemblance to Julli\`{e}re's formula, the latter takes into
account only the spin polarization of electrodes, while Eq.~(\ref{TDD})
accounts for barrier-electrode coupling and evanescent decay as well.
These contributions are the essential ones in the Fe/MgO system.

\begin{figure*}[ptb]
\includegraphics*[height=4.4cm]{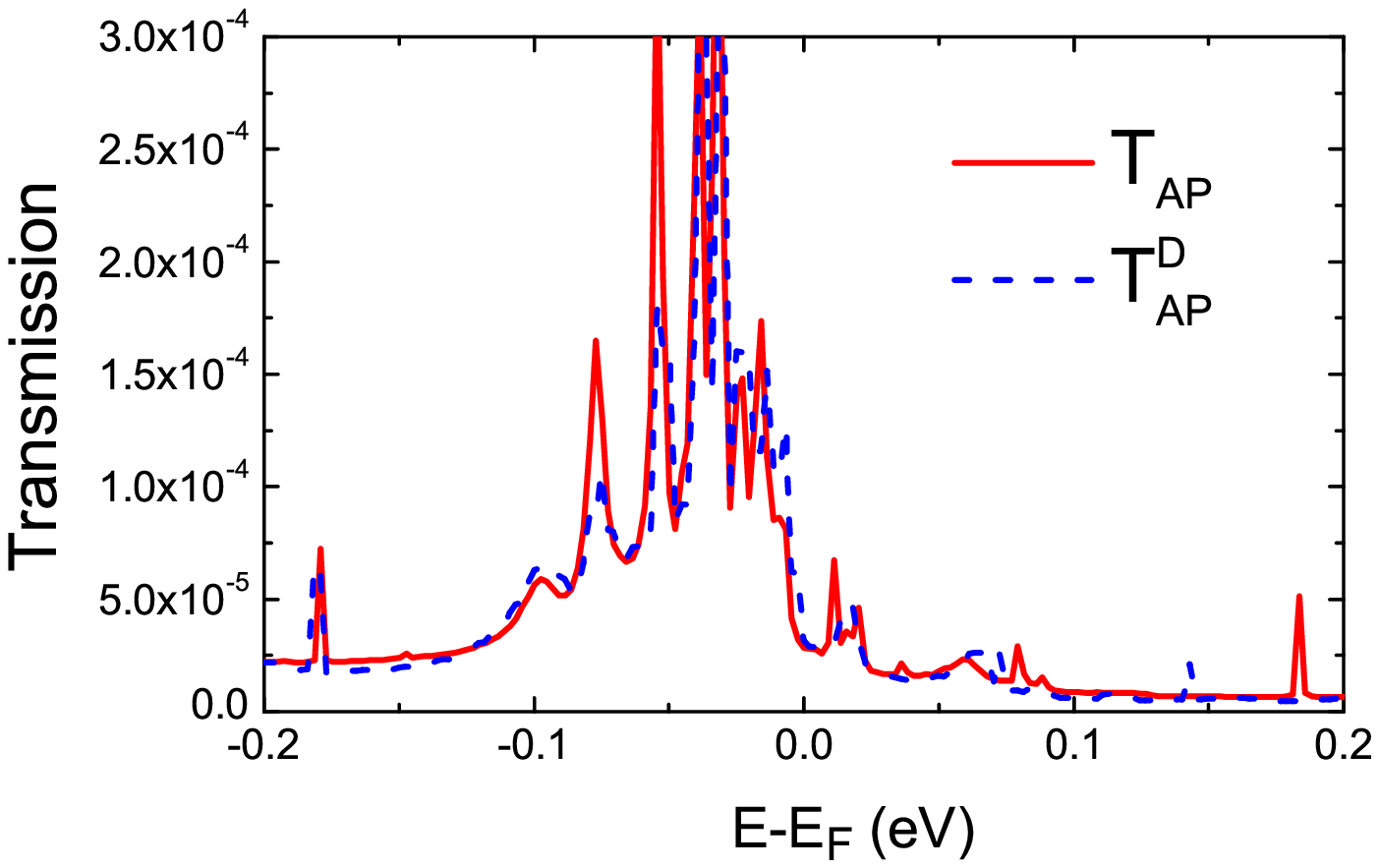}\ 
\includegraphics*[height=4.4cm]{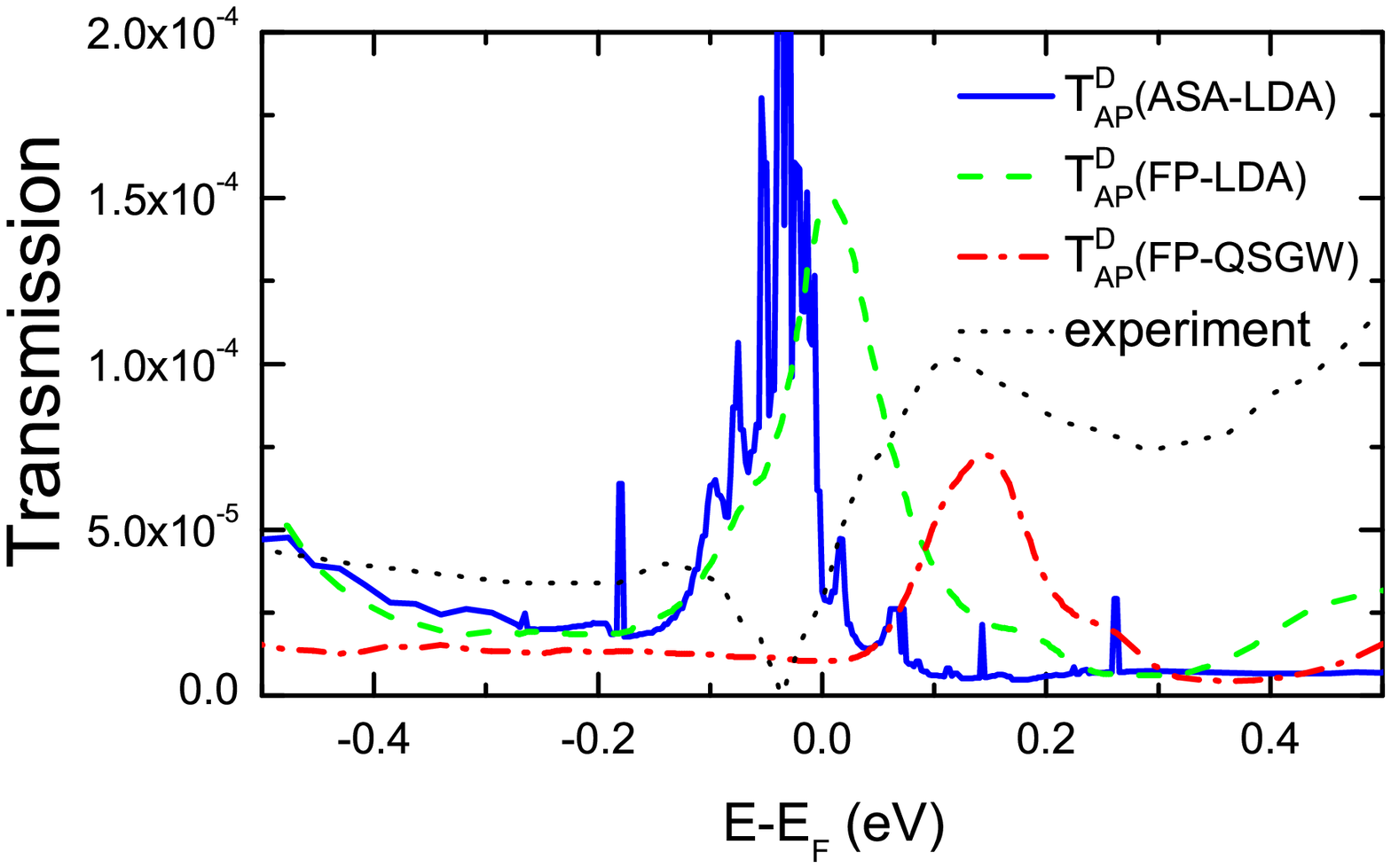}
\caption{(color online).  Transmission function in the Fe/MgO system. Left:
$T_{AP}(E)$, calculated in the APC by the TB-LMTO-ASA method for Fe/MgO/Fe
MTJ with four MgO layers (red solid line) compared to the analytic formula
$T^D_{AP}(E)$, Eq.~(\ref{TDD}) (blue dashed line), for a particular choice
of sites in the MgO where local DOS is calculated.  The latter was scaled
by $\lambda$=0.20Ry$^2$.  The quality of agreement is a measure of the
quality of approximations used to obtain Eq.~(\ref{TDD}), as described in the text.  We
verified that the close correspondence between the analytic formula and the
Landauer-Buttiker transmission seen in the left panel is insensitive to the
choice of site chosen for the local DOS.  Right: $T^D_{AP}(E)$,
Eq.~(\ref{TDD}), calculated in the antiparallel configuration for several
cases.  $T^D_{AP}$ shown in the left panel for the TB-LMTO-ASA method is
redrawn as a (blue) solid line.  The full-potential LSDA result is shown as
a (green) dashed line.
$T^D_{AP}(E)$ from \qsgw\ is shown as a (red) dash-dotted line.
Experimental data (black dotted line) shows $|d^2I/dV^2|$ measured in APC
\cite{Zermatten08} (arbitrary units).}
\label{DD}
\end{figure*}


To evaluate the trustworthiness of Eq.~(\ref{TDD}), we compare it to a
complete calculation of transmission within the Landauer-Buttiker formalism
for a system with 4 MgO layers and semi-infinite Fe electrodes.  For this
purpose we use an implementation within the tight-binding LMTO method and
Atomic Spheres Approximation (TB-LMTO-ASA)
\cite{Turek97,Schilfgaarde98}.  [As we show below the ASA is reasonably
close to, but not identical with the full potential (FP) LSDA result; it matters little
here since our purpose is to evaluate the reliability of Eq.~(\ref{TDD})].

We first calculate transmission in APC, $T_{AP}(E)$, within the
Landauer-Buttiker formalism, and compare to $T^D_{AP}(E)$ from
Eq.~(\ref{TDD}), staying within the TB-LMTO-ASA method.  For $T^D_{AP}(E)$
we employ DOS for a barrier containing 12 MgO layers (that ensures that
local DOS is converged with thickness), and use for $L$ and $R$
respectively the O atom in the third layer from the left interface
($L$=$\mathrm{O}_3^L$) and the second layer from the right interface
($R$=$\mathrm{O}_2^R$).  Thus the two DOS entering into Eq.~(\ref{TDD}) are
$\mathrm{DOS}^\mathrm{maj}_{\mathbf{k} E}(\mathrm{O}_3^L)$ and
$\mathrm{DOS}^\mathrm{min}_{\mathbf{k}
  E}(\mathrm{O}_2^R)$. $L$=$\mathrm{O}_3^L$ and $R$=$\mathrm{O}_2^R$
because oxygen atoms have more valence electrons and are located closer to
the interfacial Fe atoms; they thus better represent electrode-barrier
coupling than do the Mg atoms.  Also, for the 4-layer MgO barrier
$L$=$\mathrm{O}_3^L$ and $R$=$\mathrm{O}_2^R$ is the same atom.  Then
$d_{LR}$=0 and factor $\gamma_{\mathbf{k}E}$ in Eq.~(\ref{TDD}) is not
needed.  As Fig.~\ref{DD}$(a)$ shows, $T^D_{AP}(E)$ is
nearly identical with the Landauer-Buttiker $T_{AP}(E)$ up to normalization
$\lambda$.  We also verified that DOS taken from O atoms in other choices
of ($L$,$R$) pairs yield agreement between $T_{AP}(E)$ and $T^D_{AP}(E)$
comparable to that shown in Fig.~\ref{DD}$(a)$. Thus, the trustworthiness
of Eq.~(\ref{TDD}) is well established, and we can apply it with
justification to the FP-LSDA and \qsgw\ Hamiltonians.


The right panel of Fig.~\ref{DD} shows $T^D_{AP}(E)$, assuming the same
normalization $\lambda$ (0.20 Ry$^2$) obtained by matching $T^D_{AP}(E)$ to
$T_{AP}(E)$ in the ASA-LSDA approximation.  The ASA calculation from the
left panel is redrawn (blue solid line), and compared to a full-potential
LSDA result (green dashed line).  More precisely, the ASA approximation to
$\mathrm{DOS}^\mathrm{min}_{\mathbf{k} E}(\mathrm{O}_2^R)$ is replaced by
its analog calculated with a FP-LSDA method.
Finally we obtain $T^D_{AP}$(FP-\qsgw) from
$\mathrm{DOS}^\mathrm{min}_{\mathbf{k} E}(\mathrm{O}_2^R)$ calculated by
\qsgw\ in the repeated-slab geometry with 5 Fe and 5 MgO layers.

As Fig.~\ref{DD} shows, the peak in $T^D_{AP}$ falls near 0~V in both
the ASA-LSDA and FP-LSDA cases, in agreement with earlier full-potential
LSDA calculations \cite{Rungger09,Feng09}.
The peak of the \qsgw-derived transmission is shifted to higher energy by
approximately 0.12 eV relative to the $T^D_{AP}$(FP-LSDA) result, putting
it in close correspondance with the peak in $|d^2I/dV^2|$
(measured in APC, shown as black dashed line on Fig.~\ref{DD}.
$|d^2I/dV^2|$ is shown rather than $dI/dV$, since the features are more
easily seen \cite{Zermatten08}.) Note that
$T^D_{AP}$(\qsgw)$<${}$T^D_{AP}$(FP-LSDA).  This is because the LDA gap
(4.7 eV) is much smaller than the experimental (7.8 eV) and \qsgw (8.8 eV)
\cite{Schilfgaarde06} gaps; consequently $\gamma_{\mathbf{k}E}$ is
overestimated in the LDA.



Since finite-size effects of the leads can be important, we checked their
sensitivity by calculating $T^D_{AP}$(FP-LSDA) for slabs with 5, 7, and
9 Fe layers and 5, 7, and 9 MgO layers.  We found that the peak position
and general shape of $T^D_{AP}$ depends weakly on the number
of Fe and MgO layers.  This is because minority IRS are mostly localized
near the interface and their DOS quickly converges with number of Fe and
MgO layers.  On the other hand, finite size effects alter the majority DOS
in the repeated-slab geometry.  To eliminate finite-size effect in majority
DOS we used the TB-LMTO-ASA majority $\mathrm{DOS}^\mathrm{maj}_{\mathbf{k}
  E}(\mathrm{O}_3^L)$ obtained for the semi-infinite electrode geometry and
thick, 12-layer, MgO barrier for all three calculations.  Since ${\rm
  DOS}^\mathrm{maj}_{\mathbf{k} E}(\mathrm{O}_3^L)$ is almost independent
of energy on the scale we consider here ($E_F{\pm}$0.4~eV), the
shape and peak position of the $T^D_{AP}(E)$ will not depend on whether
ASA-LSDA, FP-LSDA, or FP-\qsgw is used to evaluate
$\mathrm{DOS}^\mathrm{maj}_{\mathbf{k} E}(\mathrm{O}_3^L)$ in the
semi-infinite limit.

\begin{figure}[t]
\includegraphics*[width=4cm,height=3.5cm]{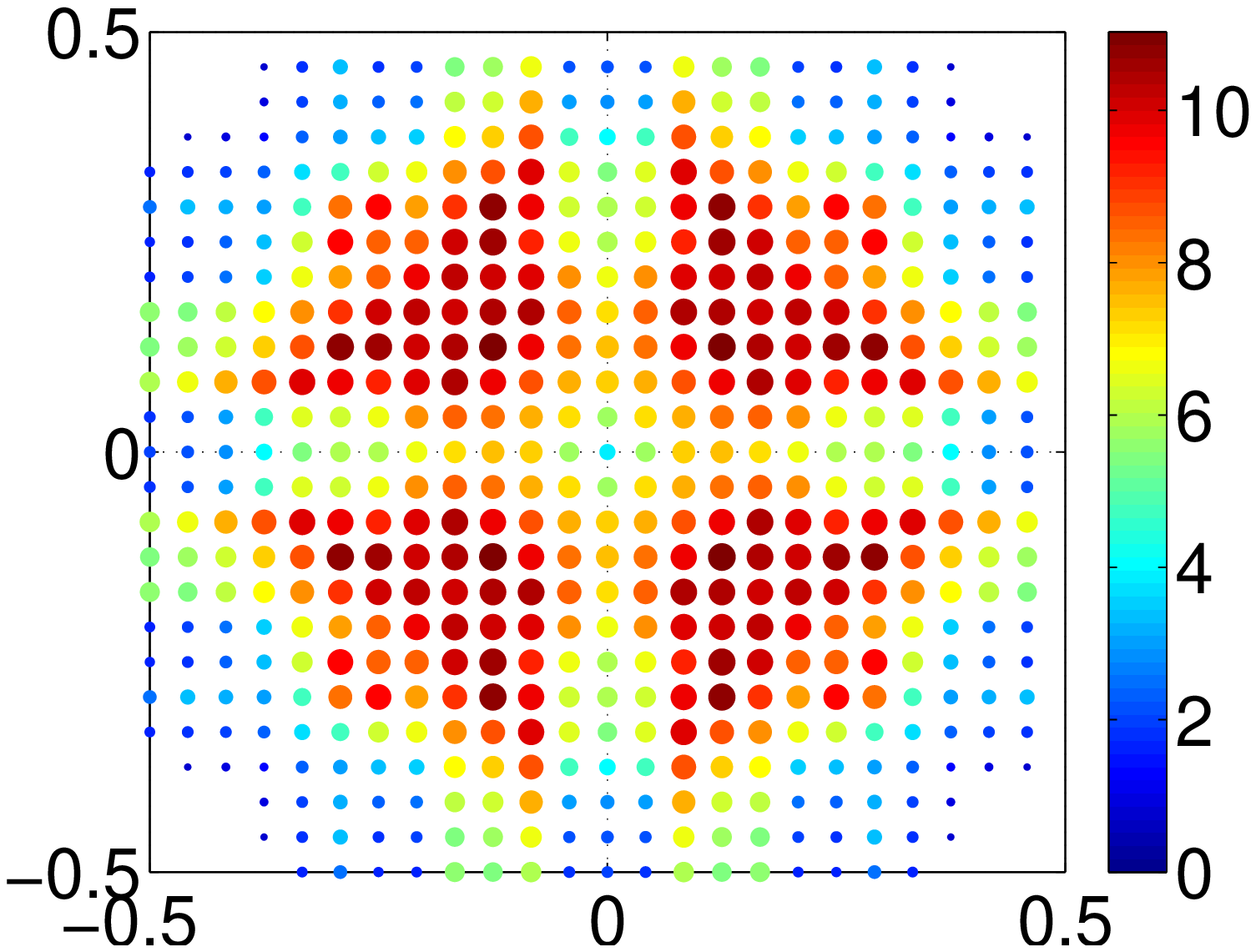} 
\includegraphics*[width=4cm,height=3.5cm]{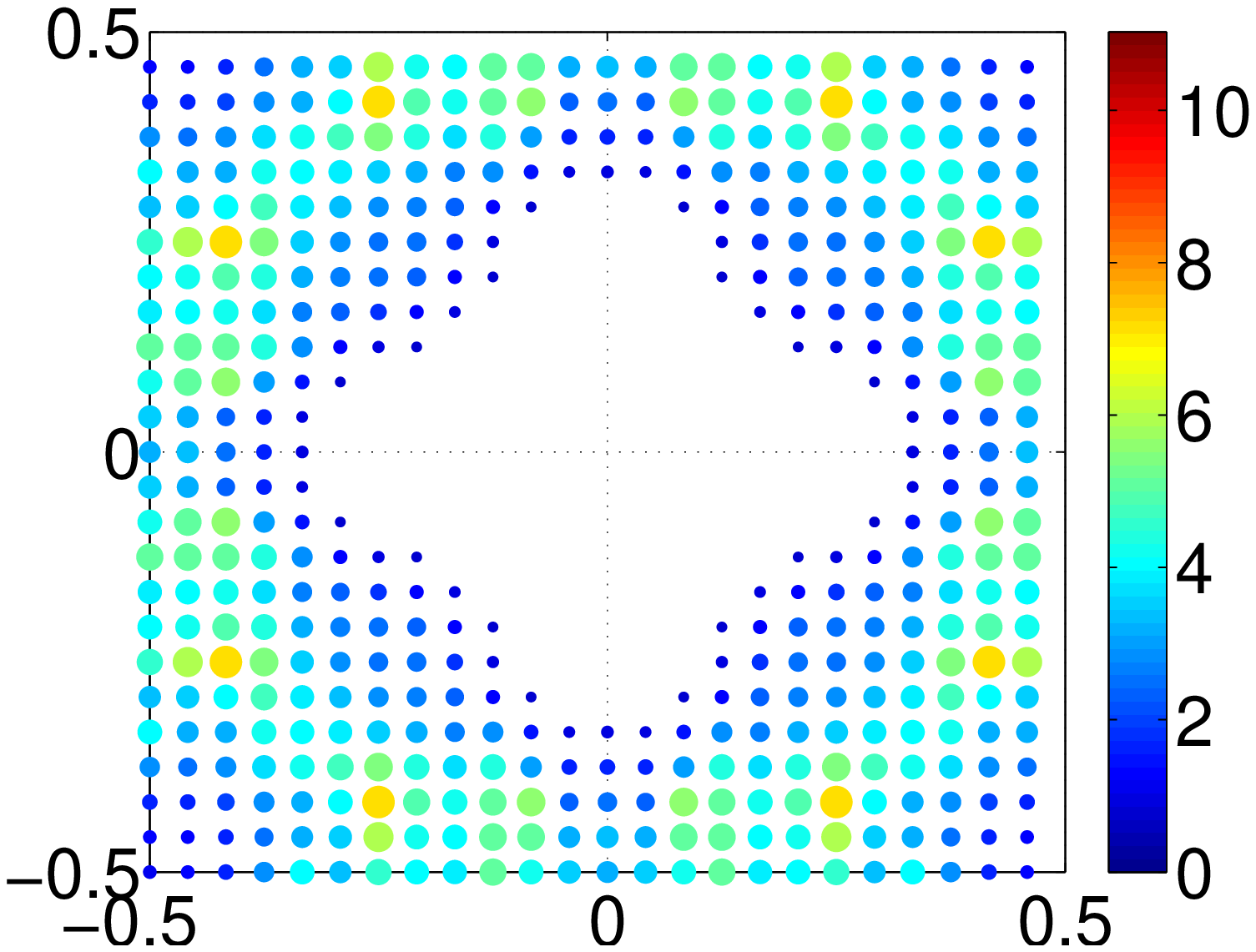}
\caption{(color online). The \textbf{k}-resolved DOS of minority electrons projected to surface Fe
layer calculated by \qsgw\  for energies corresponding to the peaks in \qsgw\  DOS, Fig. 1.
}
\label{kDOS}
\end{figure}

Significantly, there is only one peak in $T$ in both the LSDA and \qsgw\
approximations, despite the fact that two distinct peaks in the minority
DOS appear (Fig.~\ref{DOS}).  To explain why only a single peak is
seen, we analyze the \qsgw\ DOS resolved by $\mathbf{k}$ in
the 2D Brillouin zone of the (001) plane.  Fig.~\ref{kDOS} shows the
$\mathbf{k}$ resolved DOS, at $E_F+0.12$ eV and $E_F+0.26$ eV
(see two peaks in Fig.~\ref{DOS}).  As Fig.~\ref{kDOS} shows, the
$\mathbf{k}$-resolved DOS at $E_F+0.12$~eV is located mainly
around the $\Gamma$ point ($\mathbf{k}$=0), while the DOS at
the higher-energy peak ($E_F+0.26$ eV) is concentrated near the zone boundary (this feature is common to \qsgw\ and LDA).
The conduction band of MgO is free-electron like, thus  imaginary wave number
depends on $k$ approximately as $\gamma(k) \approx \sqrt{k_0^2+k^2}$ \cite{Butler01}.
Nearly all the DOS weight for the high-energy peak occurs at large $k$ where $\gamma$ is large, so its
contribution to $T$ is effectively extinguished.  This is in contrast to
the low energy peak, where the surface DOS is concentrated at small
$k$.

In conclusion, we developed a parametrization that reliably captures
the essential elements of the tunneling process in magnetic tunnel
junctions for any one-body hamiltonian.  The tunneling probability
derived from the \qsgw\ approximation are in excellent agreement with
observed differential conductance, in contrast to LSDA results.  This
confirms that the measured differential conductance peak is an
intrinsic property of the ideal Fe/MgO (001) interface.  This work
also shows that correlations treated in the \qsgw\ approximation are
sufficient to realistically describe such metal insulator interfaces.

S.F. and O.N.M acknowledge the CNMS User support by Oak Ridge National Laboratory
Division of Scientific User facilities and partial  support by  Seagate Technology.
MvS was supported by ONR contract N00014-7-1-0479 and NSF QMHP-0802216.


%


\end{document}